%% file: ms0111135rev.tex
\documentclass[12pt,preprint]{aastex}

\usepackage{psfig}

\def\spose#1{\hbox to 0pt{#1\hss}}
\def\simlt{\mathrel{\spose{\lower 3pt\hbox{$\mathchar"218$}}
     \raise 2.0pt\hbox{$\mathchar"13C$}}}
\def\simgt{\mathrel{\spose{\lower 3pt\hbox{$\mathchar"218$}}
     \raise 2.0pt\hbox{$\mathchar"13E$}}}
\def\lsim{\rlap{$<$}{\lower 1.0ex\hbox{$\sim$}}}
\def\gsim{\rlap{$>$}{\lower 1.0ex\hbox{$\sim$}}}

\def\kms{km~s$^{-1}$}

\def \deg.      {\rlap{$.$}^\circ}

\newcommand{\cms}{\mbox{cm$^{-2}$}}
\newcommand{\Lya}{\mbox{Ly$\alpha$}}

\newcommand{\hi}{\mbox{\ion{H}{1}}}

\newcommand{\alii}{\mbox{\ion{Al}{2}}}
\newcommand{\aliii}{\mbox{\ion{Al}{3}}}

\newcommand{\ci}{\mbox{\ion{C}{1}}}

\newcommand{\ciii}{\mbox{\ion{C}{3}}}
\newcommand{\civ}{\mbox{\ion{C}{4}}}

\newcommand{\feii}{\mbox{\ion{Fe}{2}}}

\newcommand{\mgi}{\mbox{\ion{Mg}{1}}}
\newcommand{\mgii}{\mbox{\ion{Mg}{2}}}

\newcommand{\mnii}{\mbox{\ion{Mn}{2}}}

\newcommand{\niii}{\mbox{\ion{Ni}{2}}}

\newcommand{\oiii}{\mbox{\ion{O}{3}}}
\newcommand{\ovi}{\mbox{\ion{O}{6}}}

\newcommand{\si}{\mbox{\ion{S}{1}}}

\newcommand{\siii}{\mbox{\ion{Si}{2}}}

\newcommand{\siiv}{\mbox{\ion{Si}{4}}}

\newcommand{\instinit}{
}
\newcommand{\instnew}[1]{
  \value{footnote}    
  \addtocounter{footnote}{-1}  
  \refstepcounter{footnote}
  \label{#1}
 }

\slugcomment{accepted by {\it The Astrophysical Journal}}

\shorttitle{Large Scale Structure at {\boldmath $z\sim 1.2$} outlined by MgII Absorbers}
\shortauthors{Williger, G. et al.}

\usepackage{natbib}
\begin{document}


\title{Large Scale Structure at {\boldmath $z=1.2$} Outlined by MgII Absorbers\altaffilmark{\ref{ack}}}


\author{Gerard M. Williger\altaffilmark{\ref{gsfc},\ref{noao},\ref{jhu}}}
\email{williger@pha.jhu.edu}
\author{Luis E. Campusano\altaffilmark{\ref{calan}}}
\email{luis@das.uchile.cl}
\author{Roger G. Clowes\altaffilmark{\ref{lancashire}}}
\email{rgclowes@uclan.ac.uk}
\author{Matthew J. Graham\altaffilmark{\ref{ucl}}}
\email{m.j.graham@ic.ac.uk}
%

\instinit
\altaffiltext{\instnew{ack}}{Based on observations with the Blanco 4m telescope at CTIO,
  which is operated by the Association of Universities for
  Research in Astronomy, Inc. under contract to the National Science Foundation.}
\altaffiltext{\instnew{gsfc}}{Laboratory for Astronomy and Solar Physics,
  NASA Goddard Space Flight Center, Code 681, Greenbelt MD~20771, USA}
\altaffiltext{\instnew{noao}}{National Optical Astronomy Observatory, 
P.O. Box 26732, Tucson AZ~85716--6732}
\altaffiltext{\instnew{jhu}}{present address: Dept. of Physics \& Astronomy, Johns 
Hopkins University, Baltimore MD 21218, USA}
\altaffiltext{\instnew{calan}}{Departamento de Astronom\'{\i}a, Universidad 
de Chile, Casilla 36-D, Santiago, Chile}
\altaffiltext{\instnew{lancashire}}
  {Centre for Astrophysics, University of Central Lancashire, Preston PR1 2HE, UK}
\altaffiltext{\instnew{ucl}}{Blackett Laboratory, Imperial College, London SW7 2BW, UK}

\begin{abstract}
The largest known structure in the high redshift universe is mapped by  at least 18
quasars and spans $\sim 5^\circ \times 2.5^\circ$ on the sky, with a quasar spatial
overdensity of 6--10 times above the mean.  This large quasar group provides an
extraordinary laboratory $\sim 100\times200\times200 h^{-3}$ comoving Mpc$^3$ in size
($q_0=0.5$, $\Lambda=0$, $H_0=100h$ \kms\ Mpc$^{-1}$) covering $1.20< z <1.39$ in
redshift.   One approach to establish how large quasar groups relate to mass
(galaxy) enhancements is to probe their gas content and distribution via background
quasars. We performed a survey for \mgii\ absorption systems in a $\sim 2.5^\circ
\times 2.5^\circ$ subfield in the large quasar group, and found 38 absorbers to a rest
equivalent width limit of $W_0=0.3$~\AA\ over $0.69<z<2.02$.  Only 24 absorbers were
expected, thus we find a 2$\sigma$ overdensity over all redshifts in our survey.   We
have found the large quasar group to be associated with 11 \mgii\ absorption systems
at $1.2<z<1.4$; 0.02\%--2.05\% of simulations with random \mgii\ redshifts match or
exceed  this number in that redshift interval,  depending on the normalization
method used. The minimal spanning tree test also supports the existence of a
structure of \mgii\ absorbers coincident with the large quasar group, and
additionally indicates a foreground structure populated by \mgii\ absorbers and
quasars at $z\sim 0.8$. Finally, we find a tendency for \mgii\ absorbers over all
redshifts in our survey
to correlate with field quasars (i.e. quasars both inside and outside of the large
quasar group) at a projected scale length on the sky of $9h^{-1}$ Mpc   and a
velocity difference $|\Delta v|=3000$ to 4500 \kms . While the correlation is on a
scale consistent with observed galaxy-AGN distributions, the nonzero velocity offset
could be due  to the ``periphery effect", in which quasars tend to populate the
outskirts of clusters of galaxies and metal absorption systems, or to peculiar
velocity effects.
\end{abstract}



\setcounter{footnote}{0}

\section{Introduction}

Evidence is mounting for the existence of super large scale structure 
on the scale of several tens of Mpc.
At low redshift ($v \simlt 40,000$ \kms ), galaxy surveys reveal
structures exceeding $\sim 50h^{-1}$ Mpc in size 
\citep[e.g.][and references therein]{Geller97,Doroshkevich00},
and
there are indications for
deviations
from homogeneity 
out to scales of 160~$h^{-1}$~Mpc \citep{Best00}.  
Structures of comparable size have
been noted in simulations of
cosmological evolution, e.g.
patterns of wall-like structure elements
with diameter 
$\sim 30-50 h^{-1}$ Mpc which surround low density regions with
typical largest extension $\sim 50-70 h^{-1}$ Mpc 
\citep{Demianski00}.
Such super large scale structure 
may be understood in
the context of the Zel'dovich nonlinear theory of gravitational instability
\citep[][and references therein]{Doroshkevich99}.
Large-box simulations can reproduce the main properties of the
observed large scale matter distribution, including structures
having a clumpy wall-like morphology, and
which incorporate $\sim 50$\% of matter with an overdensity of
$\sim 5-10$ above the mean.  
Super large scale structure thus should provide a potentially efficient
way to study large numbers of galaxies in a similar environment, 
their distribution and how they relate to phenomena such as quasars.

The most stringent constraints on models which predict the existence of
very large scale structures should be provided by measuring their properties
at early times in their evolution and thus at the highest possible redshifts.
However, at redshifts higher than a few tenths, which is the regime
accessible from large galaxy
surveys, probing
the observational characteristics of super large scale structure becomes a challenge.
One way to observe super large scale structure beyond the distances offered by large galaxy surveys
is to use brighter test objects, viz. quasars, which are much more easily detected 
than galaxies at $z>1$.  An additional advantage for tracing large structures 
offered by quasars is their intrinsic low volume density compared to galaxies,
so that a small number of them can be used to delineate structures over regions
where the expected number is on the order unity (tens of Mpc).  
The low intrinsic space density and large luminosities of quasars have already been
exploited to this end:
structures outlined by them
spanning from tens to hundreds of Mpc have been discovered at $z>1$.
These large quasar groups 
\citep[e.g.][$\sim 20$ are known to date]{Webster82,Crampton87,Crampton89,Clowes91,
Graham95,Clowes99,Clowes01}
may
represent 
high redshift precursors of large wall-like structures 
\citep[e.g.][]{Komberg94}.
Large quasar groups are not only
ideal laboratories for studying the physical characteristics of large density
perturbations in the universe, but also for 
the inter-relation between the quasars, galaxies and
gas contained within.  One difficulty is that quasars themselves trace the
highest overdensity mass perturbations, and thus are expected to be the most highly biased
tracers of mass and to cluster the most strongly, 
cf. \citet{Silk91}.  Therefore, quasars in large quasar groups 
do not by themselves reveal much about the distribution
of more common, lower mass objects in the same region.  However,
if the mass bias as a function of redshift for the quasars in large quasar groups could be
determined, then they
would provide an efficient means to
map out the large scale distribution of matter at high redshift.

The relative overdensity of various forms of matter (galaxies, gas) in large quasar groups 
(or in super large scale structure) is not
well determined.  From simulations,
\citet{Doroshkevich99}
found that from the present epoch to 
$z\sim 1$, the fraction of
matter accumulated by the largest wall-like structures for a given
density drops by a factor of $\sim 2$ and becomes negligible by $z=3$.
They suggested that
detailed statistical descriptions
of quasar absorbers are required to probe the characteristics of
super large scale structures at such epochs.  The advantage of 
quasar absorbers is that they trace much lower mass overdensities than quasars themselves,
and thus offer a much more detailed picture of the overall mass distribution.

We have used \mgii\ absorbers for just such a study, as a high redshift
pencil-beam complement to low redshift galaxy surveys.
Quasar metal absorbers delineate large structures up to 100 $h^{-1}$ Mpc
\citep{Quashnock96,Quashnock99}.
\mgii\ absorbers with rest equivalent width $W_{0,\, MgII\, 2796} \geq 0.3$~\AA\
have been strongly linked to
galaxies out to $z\sim 1.2$ 
\citep[e.g.][]{Steidel97,Guillemin97},
and thus provide a gas cross-section
selected galaxy sample, highlighting  the
cosmic web of filaments and sheets which appear to constitute
large scale structure \citep[e.g.][]{Cen97}.

An optimal region to use for a quasar--\mgii\ overdensity comparison is
the largest structure known at $z>1$.
It consists of a large quasar group of at least 18 and possibly 23 
quasars at $1.20<z<1.39$ toward ESO/SERC
field 927 
\citep{Clowes91,Clowes95,Clowes99,Newman98},
which spans $\sim 5^\circ \times 2.5^\circ$ on the sky, and has
a bright quasar space density in the region $\sim$6--10 times greater than
average.
It is ideal for study due to its 
large size and
relative proximity, thus allowing the observation of $z\sim 1.3$
associated galaxies over a range of luminosities.  
From deep optical/IR images 
in a subfield of
the large quasar group, there is evidence of a galaxy cluster
merger and 
a general excess of red galaxies around the $z\approx 1.23$ quasar J104656+0541
to a surrounding radius of $0.25^\circ$,
which are probably associated with the large quasar group \citep{Haines01a,Haines01b}.  
We have  taken spectra of
23 quasars within and behind the large quasar group, to make a survey for
\mgii\ absorbers in the region.  In the following sections, we
describe the spectra, sample selection, statistical tests to detect
overdensities and structures around the large quasar group as defined by the \mgii\
absorbers and our interpretations
of the results.

\section{Observations}
\label{sec:observations}

We obtained spectra for 23 quasars ($1.23<z_Q<2.68$)
in a $2.5^\circ \times 2.5^\circ$ field toward the Clowes \& Campusano
large quasar group with the CTIO-4m
Blanco telescope and RC spectrograph
(grating 181, Loral
3k$\times$1k CCD, 1.99~\AA /pixel, $\sim 6.7$~\AA\ resolution).
Observations were 
on the nights of 1997 March 30, April 1 and 1999 March 30,
31.  The useful wavelength coverage was 4600--9250~\AA\ ($0.64<z<2.30$ for MgII
$\lambda \lambda 2796,2804$).
Conditions were photometric with 1.1--1.8 arcsec seeing.
There were between one and seven exposures per object, with total
exposure times ranging from 900 to 10696 seconds.


The basic spectral reductions were performed with IRAF, and the 
spectra were summed using inverse variance weighting with cosmic ray
rejection routines in IDL, provided by R. Hill of the Space Telescope Imaging
Spectrograph 
group at NASA Goddard Space Flight Center.  A 1$\sigma$ error array
was
propagated throughout the reductions for each spectrum, and confirmed
with
measurements of the variation about the mean in selected
parts of the spectra.  The
wavelengths
were corrected to vacuum heliocentric values.  
Quasar positions, redshifts (based on the lowest ionization lines
available in the spectra), and photometry  \citep{Keable87,Clowes94,Clowes99}
are given in
Table~\ref{tab:observinglog}.

\section{Selection of samples}

For statistical analysis, we selected two samples of \mgii\ absorbers based on the
rest equivalent width of the $\lambda 2796$~\AA\ line.  In addition, we chose a sample
of quasars from our work and the literature, to test for correlations between \mgii\ absorbers
and quasars.  Finally, we note the existence of other metal transitions,
possible damped \Lya\ systems and the properties of a peculiar quasar in our
sample.

\subsection{Selection of \mgii\ sample}

To make a sample of \mgii\ absorption systems, we first created a
continuum for each summed spectrum with standard IRAF packages.  Next,
regions of contiguous pixels were selected for each spectrum
in which the flux
was below the continuum.  The equivalent width and corresponding error
for each region were calculated to create a list of features at 
$\geq 5.0\sigma$ significance.  We provide wavelengths of all 
features at $\geq 5.0\sigma$ significance for reference to future
higher resolution studies, to verify the existence of any metals identified
in future based on higher resolution spectra or additional data
from the \Lya\ forest.

In addition, for the sole purpose of identifying components
of \mgii\ doublets and other metal lines associated with them,
a secondary
line list of features at
$>2.5\sigma$
significance was created.
The wavelengths of the absorption features were examined for
pairs consistent with the \mgii\ $\lambda \lambda 2796, 2803$ doublet
ratio, with the requirement that the stronger $\lambda 2796$ line be
significant at $\geq 3\sigma$.  

The spectral resolution of 6.7~\AA\
is sufficient to resolve the \mgii\ doublet,
but
not to resolve blended complexes.
Therefore, as a check, each of the authors 
examined each spectrum
by eye, using velocity plots of the \mgii\ $\lambda
\lambda 2796, 2803$, to confirm the reality of each \mgii\ system.
As a further quality control measure for the doublet wavelength ratio constraint,
a majority (75\%) vote of the authors 
was required to deem a \mgii\ system as real.
We find a total initial sample of 41 ``real" \mgii\ systems (plus five candidates
and one less likely one classified as ``doubtful"), with 
the least significant ``real" \mgii\ system possessing $\sigma_{\lambda
2796}=3.36$,
$\sigma_{\lambda
2803}=2.66$ ($4.3\sigma$ summed in quadrature).  Plots of each
quasar spectrum are in Fig.~\ref{fig:rcspecplot_pub1}.
We have indicated zones in which line profiles may be 
blended with telluric absorption, determined by
1) normalizing each quasar spectrum, taking a median of the ensemble 
of normalized individual quasar
spectra and searching for features common to all spectra, and 2)
visually inspecting each spectrum to look for 
regions of absorption which are common to most or all objects.  
Expanded plots of
the
\mgii\ doublets in velocity space are in Fig.~\ref{fig:mgiivelplot}.
A list of features significant at $\geq 5.0\sigma$, plus
a small number of identified 
\mgii\ components and other metals  at $<5\sigma$ significance
associated
with \mgii\ absorbers, is in
Table~\ref{tab:linelist}.

To compare our \mgii\ absorber sample with one from
the literature,
we
define ``strong'' and ``weak'' samples with rest equivalent width thresholds
of 0.6 and 0.3~\AA , respectively, for the \mgii\ $\lambda 2796$ line 
\citep[][note that the ``weak'' sample contains
strong systems as well]{Steidel92}.
Furthermore, we restrict our analysis to 
regions of the spectrum which have sufficient signal to noise (s/n)
ratio
for each sample
such that we should be able to detect a 0.3 (0.6)~\AA\ rest equivalent width
\mgii\ $\lambda 2796$ line at $3\sigma$ significance, which requires s/n~$\simgt 18 (9)$
per 2~\AA\ pixel at 5600~\AA .  We also exclude the one absorber which lies
at a velocity separation $\Delta v < 5000$ \kms\ from its background
quasar,
as it may be an associated \mgii\ system, rather than intervening.
It was also ensured that any pairs of \mgii\ absorbers along the same
line of sight with velocity separation $\Delta v \leq 5000$ \kms\
would only be counted as one system, since there is evidence for
clustering
on that scale 
(\citeauthor{Steidel92}). 

\subsection{Selection of quasar comparison sample}
For comparison with the distribution of \mgii\ absorbers, we constructed
a sample of quasars from \citet{Veron01} and 
\citet{Keable87}, \citet{Clowes94} and \citet{Clowes99}, within a
$2.5^{\circ}$ radius of RA=10:45:00.0, dec=+05:35:00 (J2000).  There
are 107 quasars at $0.6<z<2.2$, with another 23 at $2.2<z<3.3$.  
They were found via a variety of selection methods, so we use them
as fixed reference locations for known mass concentrations in the region.
A map of the quasars in the large quasar group
and \mgii\ absorbers in the redshift range $1.20<z<1.39$ is in
Fig.~\ref{fig:quasarmap}.  


\subsection{Other metals and candidate damped systems}

We searched for other metal absorption lines by cross-correlating a list
of metal transition wavelengths with the redshifts of identified
\mgii\ systems, and found a number of \feii\ lines, as well as
\mgi , \alii , \aliii\ and \mnii .
\citet{Rao00}
found that 
approximately 50\% of the absorption systems with
$W_0({\rm MgII}\, \lambda 2796) \geq 0.5$~\AA\ and
$W_0({\rm FeII}\, \lambda 2600) \geq 0.5$~\AA\ 
have damped absorption lines meet 
the classical definition used in high-redshift surveys, 
with \hi\ column densities of
$N_{HI} > 2 \times 10^{20}$ \cms .  We have 11 absorption systems
with $W_0({\rm MgII}\, \lambda 2796) \geq 0.5$~\AA\ and
$W_0({\rm FeII}\, \lambda 2600) \geq 0.5$~\AA\ 
in our sample (Table~\ref{tab:metalsys}), three of which are within
the large quasar group.  We note that damped 
\Lya\ systems themselves often appear to indicate low mass systems
\citep[e.g.][and references therein]{Fynbo99,Warren01}, and thus
as a class may trace more typical objects in the universe than
brighter, more massive ones as quasars and Lyman break galaxies
\citep[e.g.][]{Pettini01}, though at $z>2$ the difference may be
less clear \citep[][]{Moller02}.
Nevertheless, damped \Lya\ systems
with separations of
a few Mpc may trace of large mass concentrations
at $z=2.4$ \citep{Francis00} and could prove to be similarly useful
at $z\sim 1.2$.  

\subsection{The peculiar quasar J104642+0531}

Among the objects in our sample of quasar spectra, 
J104642+0531 ($z=2.681$) is quite puzzling.  It is a peculiar background
object found serendipitously during a survey in the large quasar group field
\citep{Clowes99}. There appears to be no \civ\ emission, but there is
extremely broad \Lya\ emission ($\sim 10^4$ \kms ).  Clowes et al. found
evidence for an associated absorption system
at $z=2.654$, but no \ovi\ or other emission blueward of \Lya .
The unusual emission structure merits further study.

\section{Tests for large scale structure}

We use three methods
to test for the presence of a non-random
distribution of \mgii\ systems.  First, we calculate the redshift
distribution $dN/dz$, which will reveal whether there are any
redshift intervals with anomalously high or low counts of absorbers.
Second, we cross-correlate the quasars and \mgii\
systems in the field.  Third, we use the minimal spanning tree test
to determine whether \mgii\ systems form any connected structures.

\subsection{Redshift distribution}
\label{subsec:dndz}

To calculate the significance of any deviations of the
observed \mgii\ absorber redshift distribution, we created
control data samples which, except for clustering, accurately
reflect the 
statistical
characteristics of our data.
The specific, irregular arrangement
of detection windows in redshift space and lines of sight could create
a subtle pattern of aliasing to appear like correlations 
on certain scales, comparable to the
separation between lines of sight and the extent that each spectrum
probes along the line of sight.
To overcome these difficulties, we produced
control samples free of correlations between absorbers. 
The technique is analogous to one used in a similar study of \civ\
absorbers at $z\sim 2.4$ \citep{Williger96}, where a complete description can be found.

We used results from \citet{Steidel92},
who performed a \mgii\ survey toward 103
quasars
scattered throughout the sky to parametrize the redshift distribution
$dN/dz = N_0 (1+z)^\gamma$
where $\gamma=1.12, 1.17$ for weak, strong systems,
$W_0(\lambda 2796) \geq 0.3, 0.6$~\AA ,
and the normalization is
$\langle dN/dz \rangle = 0.97, 0.52$ at $\langle z_{abs} \rangle=1.12,
1.17$ respectively. If we integrate
over the redshift range of
each of our lines of sight to which we are sensitive to $W_0 (\lambda
2796)=0.3, 0.6$~\AA , 
\begin{equation}
N = \sum_{i=1}^{n_{quasars}} \int_{z_{low}}^{z_{high}} N_0 (1+z)^\gamma dz
\end{equation}
where the integral runs through each of the lines of sight $i$ from $z_{low}$ to $z_{high}$,
we find a total observed number of absorbers over $0.7<z<2.0$ which is 
overdense at the $\sim 2.3-2.4\sigma$ level
compared to the Steidel \& Sargent statistics 
for both the weak and strong
samples (Table~\ref{tab:absorbersample}).

We used the same parametrization to create 10000 randomized data samples, by
associating a particular sightline and redshift with a point in the
normalized cumulative redshift density function based on the actual lines of sight and
redshift limits.  The expected number of \mgii\ systems is based on a
Steidel \& Sargent's sample of 111 weak systems ($\Delta z = 114.2$) and
67 strong ones ($\Delta z=129.0$).  In comparison, our surveys represent roughly
one third of the absorbers in about one fifth of the redshift space.  The overdensity
in absorber number for our sample could be due entirely either to 
(a) an overdensity at all redshifts (cosmic variance) or
(b) the presence of one or more localized significant overdensities.
We consider each case.

To account for cosmic variance,
the number of \mgii\ absorbers in each random  sample was drawn from a Poissonian
distribution with a mean {\it equal to the number of absorbers actually
observed}.  For the strong survey, we find no significant
deviation from a random distribution 
(perhaps due to the
small sample size).  However, for the weak survey,
we find an overabundance of \mgii\ systems at 
$1.2<z<1.4$, which is coincident with
the Clowes \&
Campusano large quasar group:  we find 11 absorbers, and expect $5.5\pm 2.2$.  The
\mgii\ redshift distribution, our selection function and (for reference) the
quasar redshift distribution are shown in
Fig.~\ref{fig:dndz}a.  
If we assume a Gaussian distribution for the simulated number of absorbers at $1.2<z<1.4$,
this would be a significance of
$2.5\sigma$.  If we measure the probability directly from the number of 
simulations, 1.78\% of any of the 90000 
total redshift bins (9 bins $\times$ 10000 simulations) produced an overdensity
at the $2.5\sigma$ level;
2.05\%  had 11 or more absorbers in the $1.2<z<1.4$ bin.  
The large quasar group
occupies the  same redshift interval, which implies that the \mgii\ and
quasar overdensities are related; in fact, despite the various selection
methods used for the quasars, the ratio between quasars and \mgii\ absorbers
remains constant (within Poissonian errors of $1.1\sigma$) 
over $0.8<z<2.0$.

If there is an anomalous concentration of
absorbers
at a particular redshift, which could be the case if \mgii\ systems are
associated with the large quasar group, then the expected number
of \mgii\ systems in the same redshift 
range as the large quasar group would be overestimated
by the above procedure,
and the significance thus underestimated.  
If we draw 10000 random samples from a Poissonian number distribution
with
a mean of 24 (which is the expected number from the weak sample given
our redshift coverage and the \citeauthor{Steidel92}
redshift number density), 
then we would expect $3.4\pm 1.8$ \mgii\
absorbers at $1.2<z<1.4$, resulting in
a $4.3\sigma$ overdensity
in that redshift bin (Fig.~\ref{fig:dndz}b).  Only 0.02\%
of the simulations produced 11 absorbers in that bin, and $0.01$\% produced
more; only 0.06\% of {\it any} of the 90000 bins in any simulation had a
significance of 4.3$\sigma$ or higher.

A conservative estimate for the significance of the overdensity at
$1.2<z<1.4$ is therefore $2.5\sigma$, though it could be as high as $4.3\sigma$
depending on how we normalize our control sample.

\subsection{Quasar-\mgii\ correlations}

To test for quasar-\mgii\ correlations in three dimensions,
we calculated the three dimensional two point correlation function
between the 107 quasars 
at $0.6<z<2.2$ in our sample
and all of the \mgii\ absorbers over all redshifts in our survey.
We used 10000 control samples with randomized
\mgii\ absorber redshifts similar to those described in the previous section.
We found no
significant signal for either the strong or weak survey at any scale.

We then tested for correlations in the plane of the sky.
Although no such three dimensional large scale correlation 
has been noted in the literature, 
there is a precedent for
an association between quasars and \civ\ absorbers at a
{\it projected} distance of $\sim 10$ Mpc
(M{\o}ller 1995; private communication).  We cross-correlated the 
same quasars with the
strong and weak \mgii\
samples, but only along {\it different} lines of
sight (which avoids effects from associated absorption)
for a series of projected
separations on the sky covering
$5-50h^{-1}$ Mpc in the local frame.  Again, we used 10000 control samples
with randomized \mgii\ redshifts, drawing a number from a Poisson distribution with
a mean equal to the number of observed absorbers,
to determine the mean and standard
deviation expected in each bin.  As the number of quasar-\mgii\ pairs varied for each simulated
data set, we normalized the total number of pairs for each simulation to that actually observed.
There is no significant 
signal for the strong sample, but for the weak sample we
find a 
signal which peaks at $9h^{-1}$ proper (rest frame) Mpc projected separation
(35 arcmin at $z=1.2$)
at the $3.5\sigma$ significance level (8 pairs observed,
$2.4\pm 1.6$ expected).  
The overdensity occurs 
at a velocity difference
$\Delta v=-4500$ to $-3000$ \kms .
The negative sign indicates
that a quasar is at a lower redshift than its paired \mgii\ absorber
(Fig.~\ref{fig:quasarmgiidndv}; Table~\ref{tab:quasarmgiipairs}).  
We expect the peak to be 
at $\Delta v \sim 0$ if the quasar redshifts are accurate, and
if quasars and \mgii\ systems trace mass in a similar way.
Only 0.23\% of the
simulations produced as many as 8 pairs in the
$\Delta v=-4500$ to $-3000$ \kms\
velocity bin, with 0.26\% of
{\it any} bins among all of the Monte Carlo simulations producing an equal
or greater overdensity of $\geq 3.5\sigma$.  
There is no significant preference for strong or weak systems to be associated
with the correlation.   
Most of the signal (i.e. 5 of 8 pairs)
comes
from quasars and \mgii\ absorbers in the large quasar group.  Possible physical
explanations for the overdensity will be discussed in \S\ref{sec:discussion}.

It is possible, but unlikely, that the quasar-\mgii\ absorber pair
velocity difference is produced by a systematic quasar offset
between quasar rest frame UV and optical lines.
Our quasar redshifts were taken either from 
the \citet{Veron01}
catalogue,
measured from our own
$\sim 10$~\AA\ confirmation spectra \citep{Clowes94, Clowes99}
or, in the case of 22 of the
quasars (all except J104545+0523), from the data presented here.
There is no systematic offset between the our two sets of measurements,
independent of whether \mgii , \civ , \siiv\ or \ciii ] emission lines were used.
Five of the eight quasar-\mgii\ pairs at $-4500 < \Delta v < -3000$ \kms\
have quasar redshifts determined from \mgii\ emission lines.  The other
three are from higher ionization \civ\ or \ciii ] emission lines.
\citet{McIntosh99} and \citet[][and references therein]{Scott00}
find that
quasar \mgii\ emission  provides redshifts within
$\sim 400$ \kms\ of [\oiii ] 5007 (the systemic
redshift fiducial), with a correlation between velocity differences and
quasar luminosity.  
However, the observed 
peak in the distribution of quasar-\mgii\ absorber pairs
is at $\Delta v\sim 10$ times larger than would be expected from 
an offset between \mgii\ (rest frame UV) and the fiducial 
[\oiii ] 5007 (rest frame optical) emission.  IR spectroscopy of the quasars
which produce the quasar-\mgii\ pair overdensity can confirm or rule out this
rather unlikely explanation.

The luminosities of the quasars involved in the correlation are not unusual.
The 8 quasars in the
overdensity of pairs with $\Delta v=-4500$ to $-3000$ \kms\
have mean absolute magnitude $\langle M_{abs}\rangle = -25.8\pm1.0$,
calculated with code kindly provided by M. Veron-Cetty for the purposes of
using a uniform definition of $M_{abs}$.  Our sample of
quasars at $0.6<z<2.2$ 
has $\langle M_{abs}\rangle = -25.7\pm 1.0$,
whereas 16892 quasars in the \citeauthor{Veron01}
catalogue at
the same $z$ range have $\langle M_{abs}\rangle = -25.4\pm 2.4$.
Our quasar sample is  not significantly more luminous
than a very large but admittedly inhomogeneous sample, so it is doubtful
that a luminosity effect 
contributes to part of the systematic
velocity difference we observe between the quasars and \mgii\ absorbers.
Indeed, the correlation may simply be a fluke due to small number
statistics, and should be confirmed with a larger data sample.

\setcounter{footnote}{0}

\subsection{Minimal spanning tree}

The minimal spanning tree (MST) is a heuristic algorithm which can
delineate and characterize structure within a data set. We have applied
the technique of \citet{Graham95}
to search for clusters of \mgii\
absorbers in both the strong and weak samples. An identified cluster is
assigned a statistical significance by determining how frequently
structures of equivalent 
multiplicity\footnote{Multiplicity is a term used in spatial pattern 
statistics to denote the
number of objects making up a cluster.}
and with a more clustered
morphology occur in 10000 simulations of the data. 
A simulated absorber position is 
assigned the right ascension and declination of a randomly selected real absorber to
maintain any selection effects in the plane of the sky. Its 
redshift is determined according to one of three prescriptions:
(1) drawn from the observed redshift distribution binned in bins of width
$\Delta z = 0.2, 0.4$; 
(2) drawn from the \citet{Steidel92} redshift
distribution; and (3) as (2) but with the total number count normalized to
the observed number count. For (1), the bins were selected either to match
the large quasar group width or to double it, to smooth out Poisson fluctuations on
scales of half a bin width or smaller.
For (2) and (3), cosmic variance is also taken
into account as previously described. 
No significant structures are found in the strong survey, but a cluster of
10 absorbers is found in the weak survey at redshift $z \sim 1.3$.
The significance level for the $z\sim 1.3$ structure 
from prescription (1) is $P=0.005$ (0.002) for
$\Delta z = 0.2$ (0.4); for prescriptions (2) and (3), $P=0.04$, 0.01
respectively.

{\it Foreground structure at $z\approx 0.8$.}
At lower redshift, 
the MST test shows a group of 7 absorbers at $0.77<z<0.89$
with probability to reject that the decision to reject 
the null hypothesis (that the absorbers are
intrinsically unclustered) is wrong
of
$P\approx 0.005-0.02$. The exact value of $P$ depends on which of the three prescriptions
in the preceding paragraph was used to create the control samples.
$P$ is largely independent of the choice of redshift bin width for the controls
($\Delta z = 0.1, 0.2,
0.4$).
It appears that at least one
of the absorbers is very close (within 1 arcmin) to a galaxy cluster at $z\approx 0.8$
\citep{Haines01a}.  There is also a structure of 14 quasars at
$z\approx 0.8$, which is consistent with a random distribution with probability
$P_{config}=0.066$ for the observed multiplicity, 
as determined from a MST analysis of the Chile-UK Quasar Survey \citep[CUQS,][]{Newmanthesis}.
The probability is again independent of control redshift bin widths.
The number of quasars from our sample \citep{Newmanthesis} is lower at $z<1$,
with the probability of any structure arising at $z<1$ of
$P_{z<1}=0.14-0.19$.  Thus, the probability
of seeing the observed structure at $z\approx 0.8$ is $P_{config}P_{z<1}\sim 0.01$.
The coincidence of the \mgii\
absorber group, the quasar group and its size, and the proximity of a
cluster of red galaxies support the notion of a foreground structure.

\section{Discussion}
\label{sec:discussion}

The coincidence of the MgII absorber candidate overabundance with the
large quasar group implies that the large quasar group is accompanied by a corresponding increase in
galaxy density, possibly
similar to those found associated
with multiple \civ\ systems at $z\sim 2$ by 
\citet{Aragon94}.  The MST result
provides an
independent test which supports the existence of 
a structure of \mgii\ absorbers within the
large quasar group.  
The quasar-\mgii\ correlation may reflect a characteristic size of filaments.

\subsection{\mgii\ absorber overdensity}

It may be that \mgii\ absorbers within the large quasar group could
be associated with nearby enhanced ionizing sources such as 
undetected quasars or AGN.
In that case, the halos of the galaxies responsible for the
\mgii\ absorption could possess smaller detectable gas cross
sections than in the field, the ``galaxy proximity effect"
\citep{Pascarelle01}, which would cause us to underestimate the
significance of the \mgii\ absorber (galaxy) number density in the region.  
Any mass estimates for the region would also have to take into account that
numerical simulations and semi-analytic models show that galaxies
should be more highly biased tracers of the mass at higher redshift.
If so,
then the overdensity of matter in galaxies which produce the \mgii\ absorption
could be closer to the overdensity of matter in super large scale structure of $5-10$ above the mean predicted by 
\citet{Doroshkevich99}.
Estimates of the matter associated with \mgii\ absorber overdensities
could be redshift-dependent:
in CDM models,
as the amplitude of galaxy clustering remains roughly fixed, the dark
matter structure grows with time.   
Large quasar groups and the galaxies they contain could provide the means to test for
such a trend:
direct imaging of the \mgii\ absorbers in the large quasar group should
reveal whether there is a tendency for them to occur in areas of
higher than average galaxy density, and velocity dispersions of 
any associated galaxy clusters/groups should constrain the amount of matter in the
vicinity.  High resolution imaging of \mgii\ absorbers in the large quasar group 
could also reveal whether quasars behind the large quasar group are being
lensed by galaxies within the large quasar group, which would make it more likely to observe
bright background quasars in the region and find foreground \mgii .

Though rare,  large structures at $z\sim 1$ have clearly been noted in
simulations. \citet[][Hubble volume simulations]{Evrard01}
find a large
cluster at $z=1.04$ in a $\Lambda$CDM model which has a mass twice that of Coma, a
line-of-sight velocity dispersion of  $\Delta v = 1900$ \kms\ and an equivalent
X-ray temperature of 17 keV.  It is larger than any known cluster.  Although
unusual, it might be representative of parts of
super large scale structure at
$z\sim 1$.  
X-ray observations of our large quasar group field, for example around the merging galaxy
clusters 
imaged in the optical and near IR by \cite{Haines01b}, would 
provide the most direct confirmation for such large 
density perturbations.
If baryonic gas evolution can be linked to the dark matter halos in
simulations such as that of the Virgo Consortium (which is admittedly 
difficult, as
\mgii\ arises in galaxies), it could be possible to make mock
pencil-beam surveys through the simulated data, and thus make a direct comparison between
the observed and predicted  size and frequency of large quasar groups and the  distribution of
galaxies/\mgii\ absorbers within them.  However, it would be more feasible to
study Hubble volume simulations of
\Lya\ absorbers (which do not necessarily arise in galaxies), and to compare their
distribution to
the presence
of very large structures.  In that case, a comparison with observations 
would require HST spectroscopy for our target field.  

A possible non-gravitational origin of such large structures as large quasar groups could be
the destruction of H$_2$ or the ionization of He, both of which can be
affected over one to several tens Mpc by luminous quasars or other effects of
``cooperative galaxy formation"
\citep{Miralda00,Ferrara98,Bower93,Kang92,Babul91}.
However, enhanced
photoionization has been proposed to {\it impede}\, galaxy formation, at least
for low mass halos,  by
heating the intergalactic medium, inhibiting the collapse of gas into dark
halos and reducing the radiative cooling of gas  within halos, though
galaxies in deep potential wells (brighter than $L_*$)  appear unaffected
\citep{Benson01}.  Large scale perturbations approaching large quasar group size can produce
bias on similar size scales, effectively reducing the galaxy formation
efficiency in surrounding lower density regions, and thus enhancing the
contrast of very large scale structure \citep{Demianski99}. The effect of
inhomogeneous photoionization on structure formation clearly deserves further
investigation.

\subsection{Quasar--\mgii\ absorber correlation}

The quasar-\mgii\ absorber correlation may reflect the size of LSS filaments, and
is close to the 7~Mpc size of filaments associated with low $z$ \Lya\ forest clouds
from simulations of structure evolution \citep*{Petitjean95}. The angular scale
where the peak number of pairs is observed (corresponding to  35 arcmin at
$z\approx 1.2$) is consistent with the correlations of up to $1^\circ$ between
$-24<M_B<-19$ AGN and early-type galaxies detected by  \citet{Brown01}.   We note
that five of the eight quasar-\mgii\ system pairs occur within the large quasar group
($1.24<z<1.35$). The five large quasar group quasar-\mgii\ absorber pairs as an ensemble produce
most of the significance in the distribution of pairs as a function of velocity (an
overdensity  at the $\sim 2\sigma$ level), and form  a pair and a triplet of
quasars of scale 1-10$h^{-1}$ Mpc, a size on the order of that expected for
filamentary structure. The peak at $-4500 < \Delta v < -3000$ \kms\ should be
confirmed with IR spectroscopy of the quasars in question, to rule out the unlikely
possibility that the large velocity difference arises from high {\it vs.} low
ionization line emission regions. Otherwise, the quasars and MgII systems could
trace different parts of the same filamentary substructure within the large quasar group.  Such a
geometry may result as a consequence of the ``periphery effect", in which quasars
tend to populate the peripheries of galaxy clusters and metal absorption line
groups \citep{Jakobsen92,Sanchez99,Tanaka00,Haines01b,Sochting02}.   In a
particularly analogous situation to the large quasar group in this study, \citet{Tanaka01}
detect clustering of faint  red galaxies ($I>21$, $R-I>1.2$) over a scale
extending to 10~$h^{-1}$ Mpc around a tight group of five radio-quiet quasars at
$z\sim 1.1$, which are embedded in  the \citet{Crampton87,Crampton89} large quasar group of 23 quasars.
\citet{Haines01b} report a similar phenomenon around  the $z\approx 1.23$ quasar
J104656+0541 within the large quasar group studied here, albeit in a smaller observed field.

If the periphery effect is the cause of our observed quasar-\mgii\ correlation at
non-zero velocity difference, then we should in principle detect a signal at both
positive and negative $\Delta v$ (according to our definition). However, we only
see a negative velocity peak, in which the largest fraction of the signal arises
from a set of quasars at $z=1.236, 1.273, 1.316$  separated by $36.9
\times 34.3 \times 52.2$ arcmin and 4900, 5600 \kms\ in front of a  trio of \mgii\
absorbers separated by 5100, 5700 \kms\ toward two sightlines 29.8 arcmin apart. 
If the correlation is real, in larger data samples  we would expect to find cases
of quasars both in front of and behind groups of \mgii\ absorbers. Alternatively,
such a quasar-\mgii\  correlation which arises for projected distances on the sky,
but not in three dimensions, could arise as an effect of peculiar velocities along
the line of sight, and may be related to the  ``bull's-eye effect''
\citep{Praton97,Melott98}, in which peculiar velocities in collapsing structures
tend to make structures more significant in redshift space than in real space.  
The number of quasar-\mgii\ pairs in the large quasar group is not sufficient to produce a
significant signal on its own, however, and the correlation should be confirmed
with a larger sample.

If the quasar-\mgii\ projected correlation is found to be real, it could
be a useful tool to probe the extent and evolution of the 
overdensities giving rise to quasars themselves.
It is possible that quasar clustering
declines toward low redshift, as quasar activity moves from the most
massive galaxies into lower mass systems that are less highly biased.
We expect to continue to find more members of the large quasar group toward
ESO/SERC field 927,
which will provide better statistics
with which to study the  relationship between 
various matter overdensities in the form of
quasars, \mgii\ systems and galaxies.

%



\acknowledgments 

We thank Romeel Dav\'e, Andrei Doroshkevich,
Gus Evrard, Dick Hunstead, Palle M{\o}ller, Martin Rees, Steve Warren,
David Weinberg and Simon White for useful discussions and comments on a draft of
this paper,
Robert Hill for data reduction software and Alvaro Amigo for assistance with
the data reduction.  We also thank the anonymous referee for helpful
suggestions concerning the clarity of our presentation and results.

\clearpage
Fig.\ref{fig:rcspecplot_pub1}: (a-f) Spectra in the quasar sample.  The $1\sigma$ error
  is shown by the solid line close to the bottom of each plot.
  The dashed and dotted lines show the wavelengths used for the strong
  and weak \mgii\ surveys, respectively.  Regions deemed affected by telluric absorption are shaded.
Absorption lines are indicated by
  ticks above the spectrum at $\geq 5\sigma$ (solid) and $2.5<\sigma <5$
  (dashed) significance; in the telluric bands, 
  only identified extragalactic features are ticked.  However, all absorption
  features are listed in Table~\ref{tab:linelist}.
\begin{figure}[ht] 
\begin{center}
  \caption[]{
    \label{fig:rcspecplot_pub1}
    }
\end{center}
\end{figure}

\clearpage

Fig.\ref{fig:mgiivelplot}: (a-d) Plots of the \mgii\ doublets in velocity
  space.  Normalized fluxes are shown.  The $\lambda$ 2803 component is shown
  offset by -0.2 in flux for clarity.  The $1\sigma$ error array is indicated
  by the dot-dashed line. The background quasar is listed to the bottom left of
  each plot, the \mgii\ absorption redshift is at the bottom center, and four
  letters indicating the strength of the line (s=strong, w=weak, v=very weak),
  whether the absorber is in the strong and weak surveys (y/n) and the reality
  of the system (r=real, c=candidate, d=doubtful), as listed in Table~\ref{tab:metalsys}.
  Only ``real" systems which are in the strong or weak surveys are used in the
  analysis for this work.

\setcounter{figure}{1}
\clearpage
\begin{figure}[ht] 
\begin{center}
  \plotone{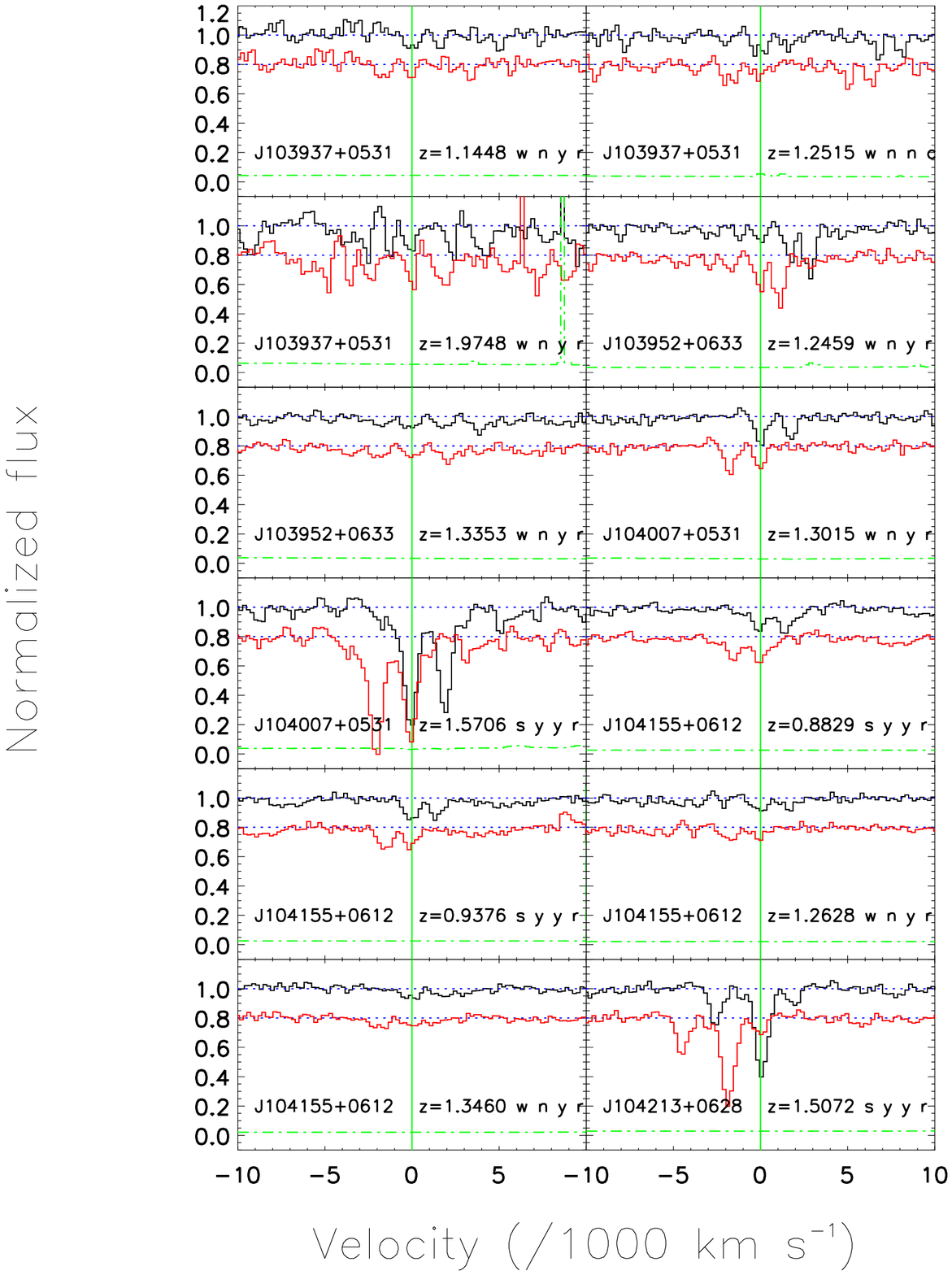} 
  \caption[]{(a)
    \label{fig:mgiivelplot}
    }
\end{center}
\end{figure}

\setcounter{figure}{1}
\clearpage
\begin{figure}[ht] 
\begin{center}
  \plotone{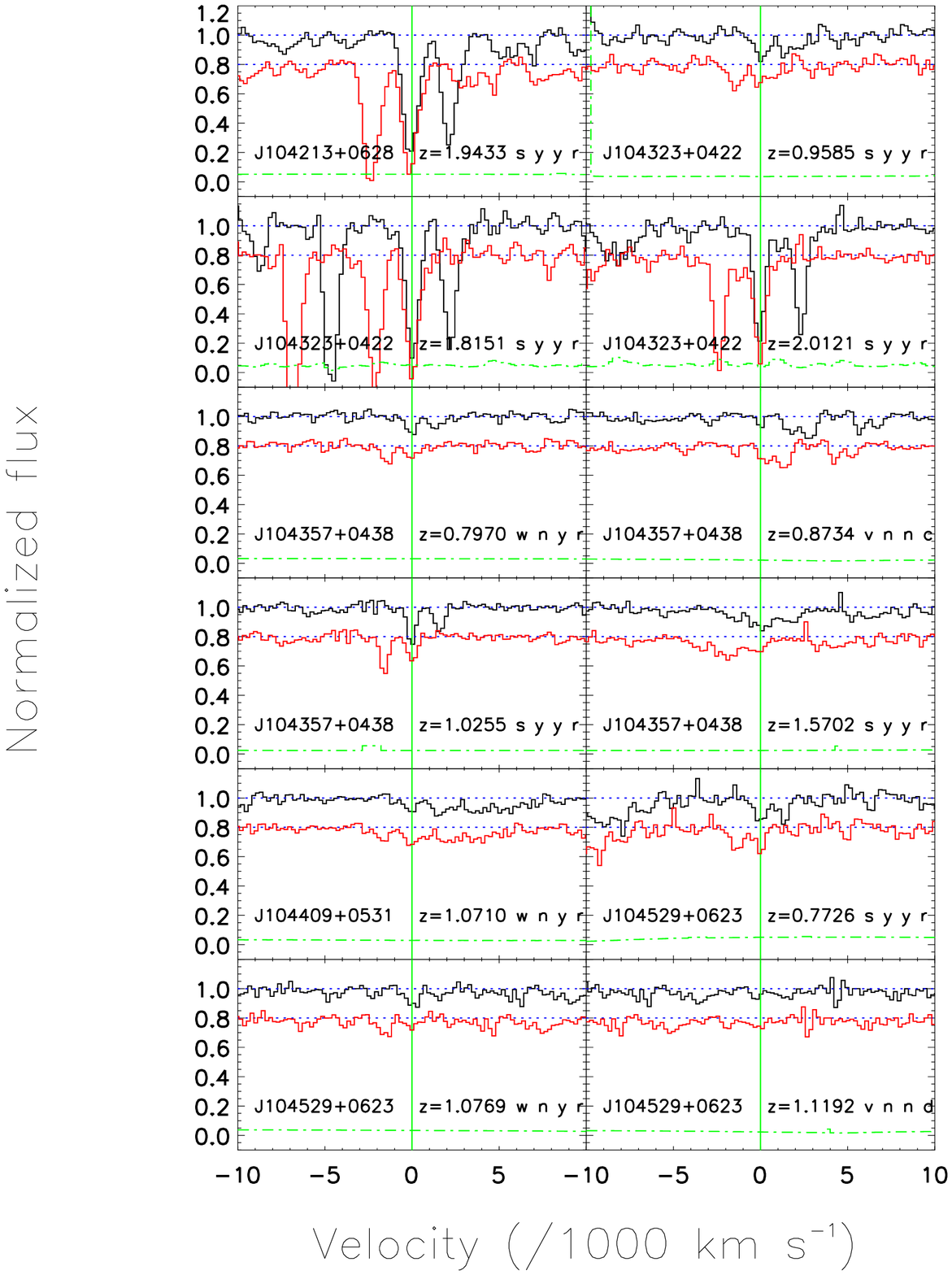} 
  \caption[]{(b)
    }
\end{center}
\end{figure}

\setcounter{figure}{1}
\clearpage
\begin{figure}[ht] 
\begin{center}
  \plotone{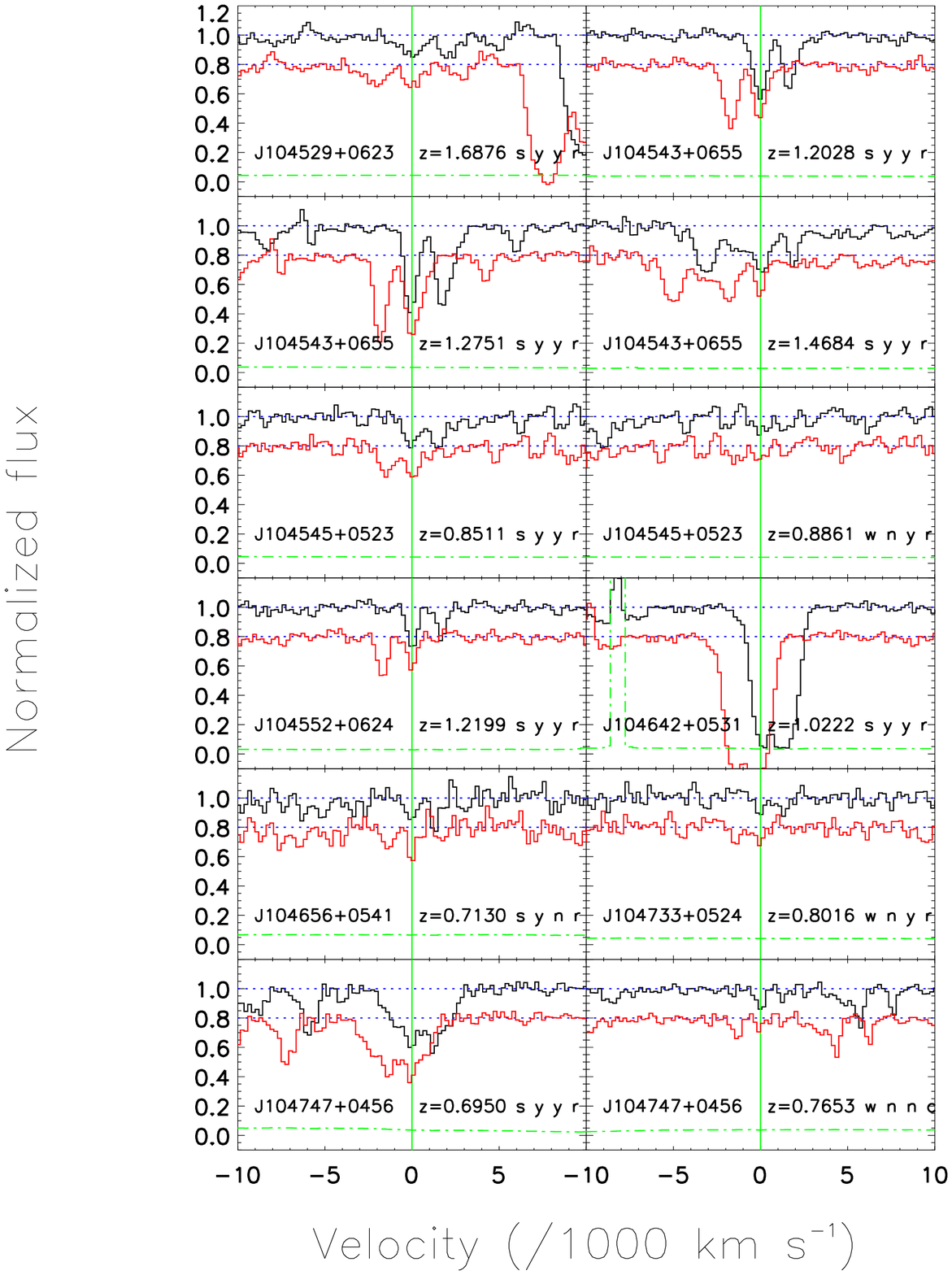} 
  \caption[]{(c)
    }
\end{center}
\end{figure}

\setcounter{figure}{1}
\clearpage
\begin{figure}[ht] 
\begin{center}
  \plotone{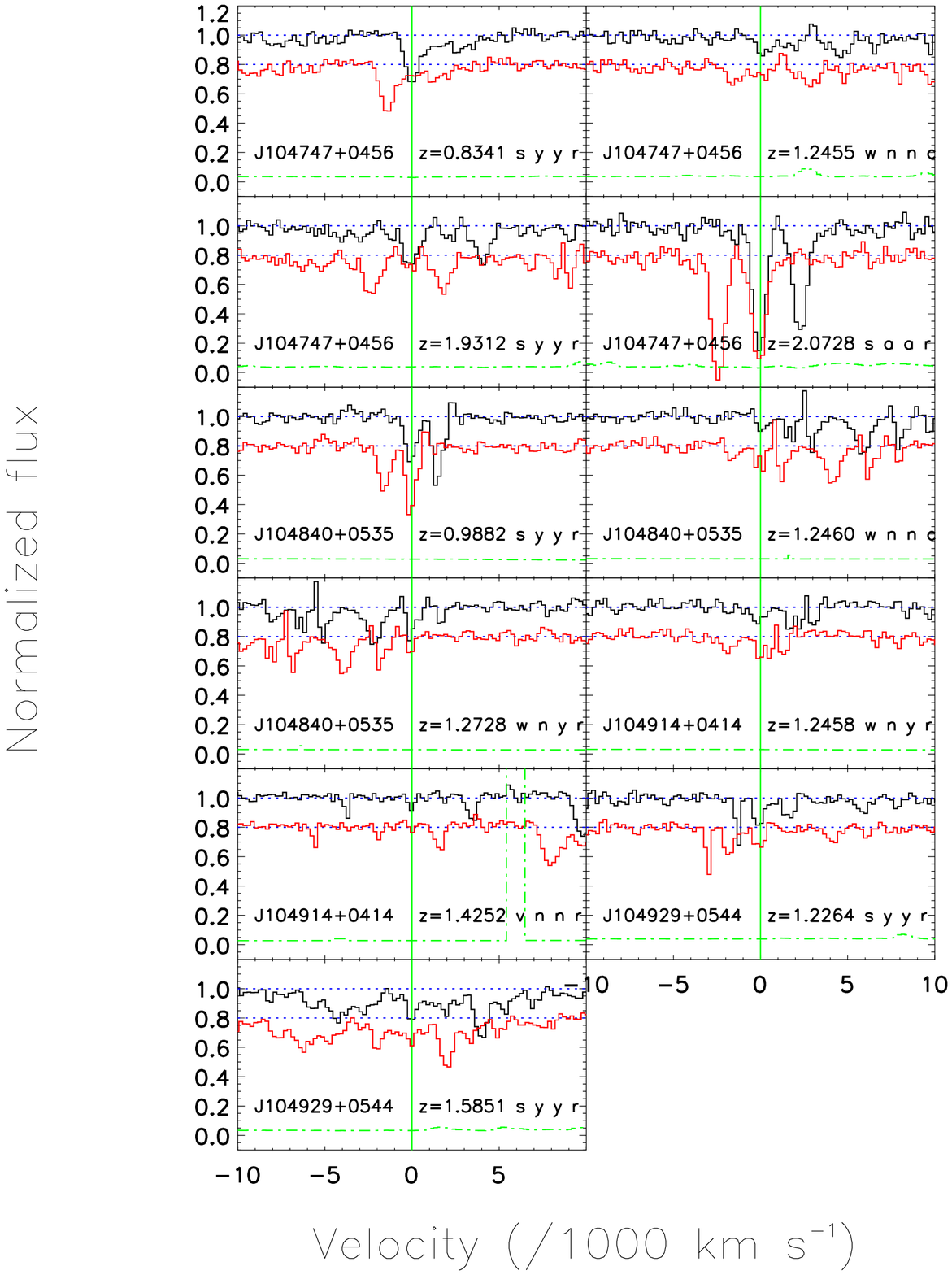} 
  \caption[]{(d)
    }
\end{center}
\end{figure}

\clearpage

\begin{figure}[ht] 
\begin{center}
  \plotone{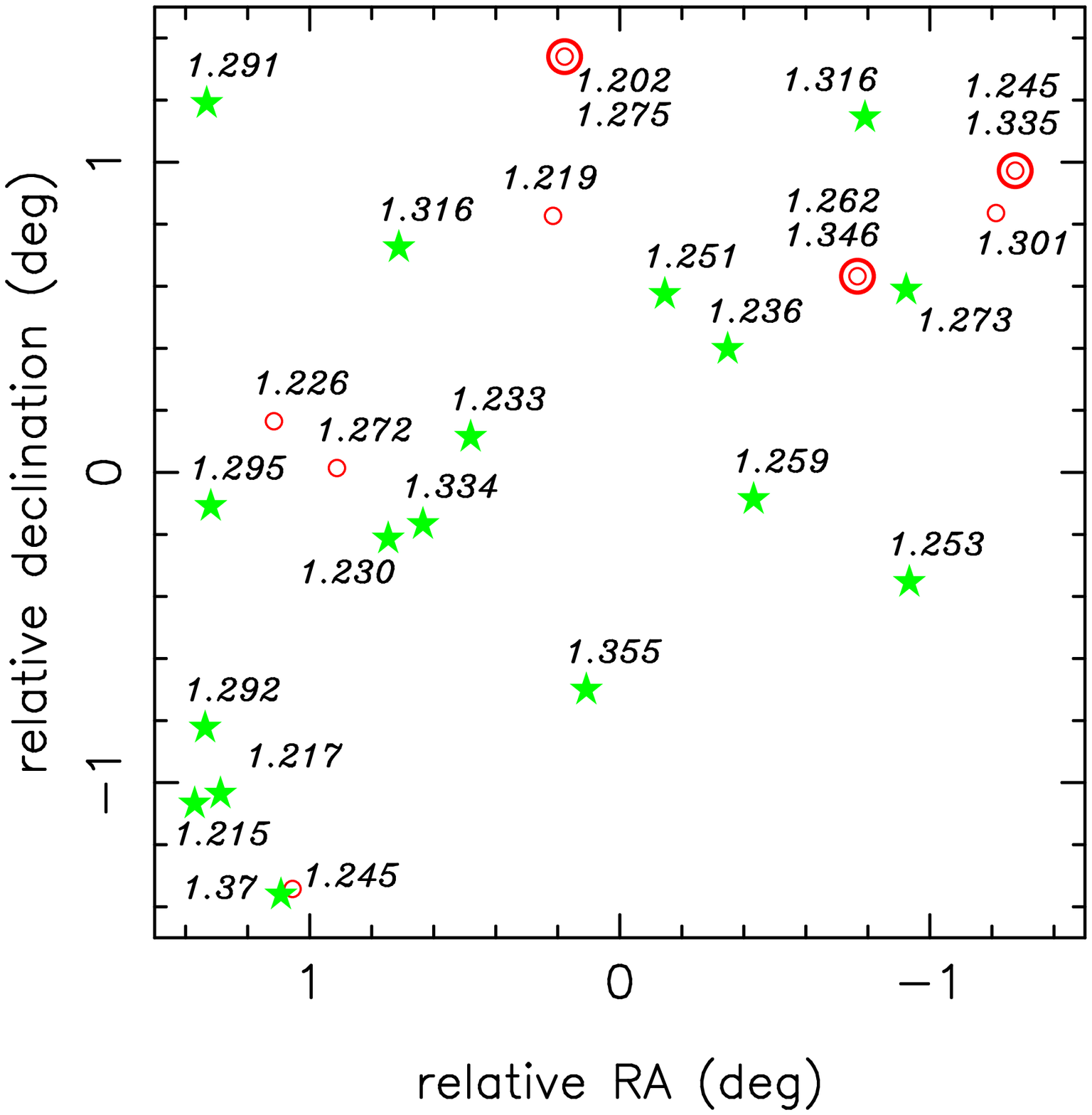} 
  \caption[]{Map of the quasars (stars) and \mgii\ absorbers 
(circles),  at $1.20<z<1.39$ toward the Clowes \& Campusano large quasar group.
{\it Stars}\, indicate positions of QSOs in the redshift range $1.20<z<1.39$.
{\it Circles}\, indicate the positions of quasars with $z > 1.39$
with \mgii\ absorption at $1.20<z<1.39$.
Double circles indicate two \mgii\ systems toward a line of sight.
Redshifts for the quasars (stars) and \mgii\ absorbers (circles)
are listed adjacent to each symbol.  The field center is at
RA=10:45:00.0, dec=+05:35:00 (J2000). About half of the structure is shown. 
    \label{fig:quasarmap}
    }
\end{center}
\end{figure}
\clearpage

\begin{figure}[ht] 
\begin{center}
  \plotone{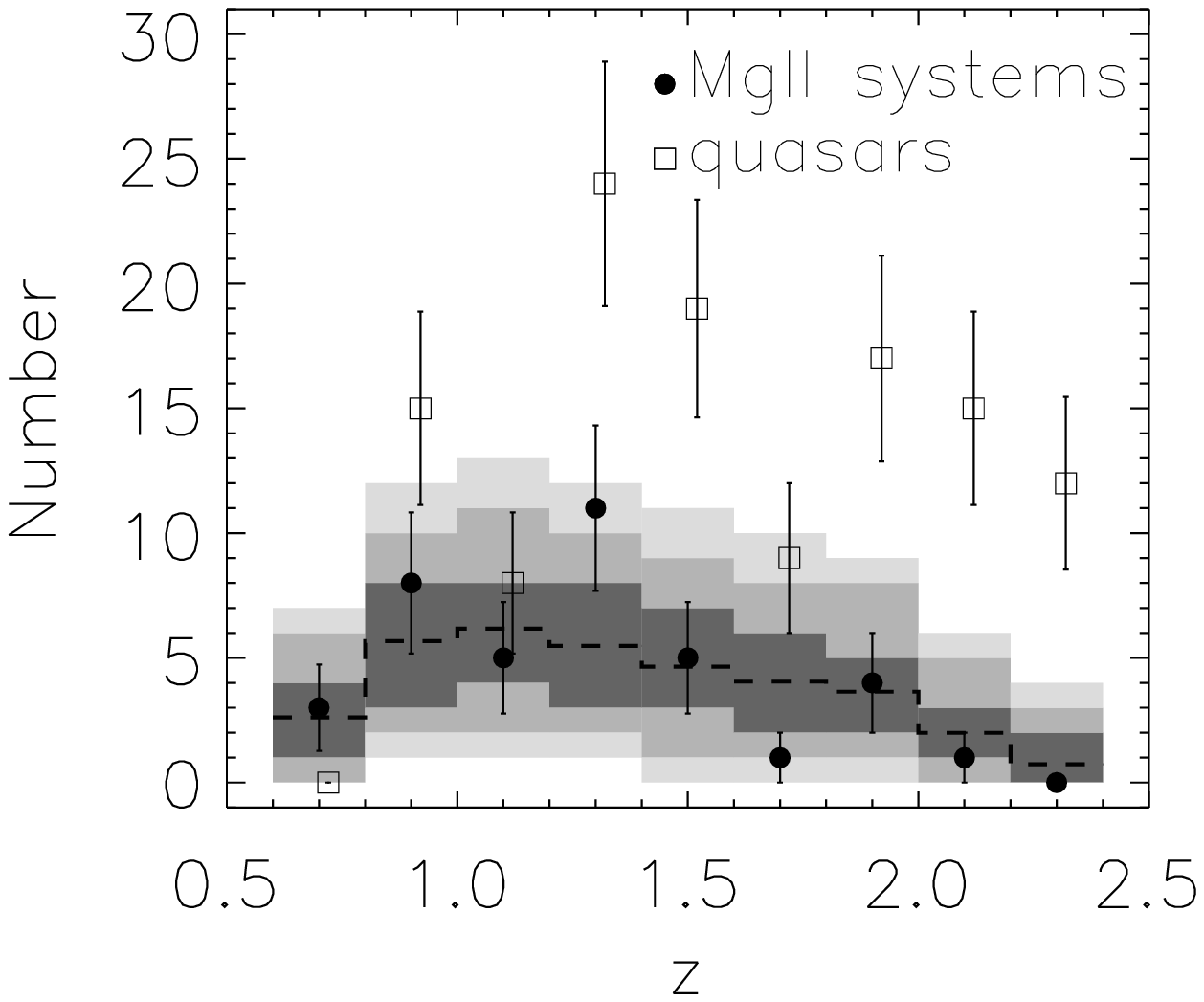} 
  \caption[]{(a) 
    \label{fig:dndz}
    }
\end{center}
\end{figure}

Fig.\ref{fig:dndz}: 
(a) Redshift distribution of the weak \mgii\ absorber
  sample, compared to simulations of the number \underline{observed}.   
{\it Filled
    circles:} observed, with errors assuming a Poissonian distribution. 
    {\it Dashed line:} mean expected number per bin from 10000
  Monte Carlo simulations, drawn from samples with mean totals equal to the 
  number observed (38).
  {\it Shaded regions:} 68, 95, 99\%  scatter about
  the expected mean.  The overdensity at $1.2<z<1.4$, which coincides with
  the large quasar group, was matched or exceeded
  in 2.05\% of the simulations.  {\it Open boxes (slightly offset in $z$ for clarity)}: known quasars in
  the field, with errors assuming a Poissonian distribution.\\
  (b) Redshift distribution of the weak \mgii\ absorber
  sample, compared to simulations of the number \underline{expected}.  
{\it Filled
    circles:} observed, with errors assuming a Poissonian distribution. 
    {\it Dashed line:} mean expected number per bin from 10000
  Monte Carlo simulations, drawn from samples with mean totals equal to the number 
  expected (24).
  {\it Shaded regions:} 68, 95, 99\%  scatter about
  the expected mean.  The overdensity at $1.2<z<1.4$ was matched or exceeded
  in 0.02\% of the simulations.  {\it Open boxes (slightly offset in $z$ for clarity):} 
  known quasars in the field, 
  with errors assuming a Poissonian distribution.

\clearpage

\setcounter{figure}{3}
\clearpage
\begin{figure}[ht] 
\begin{center}
  \plotone{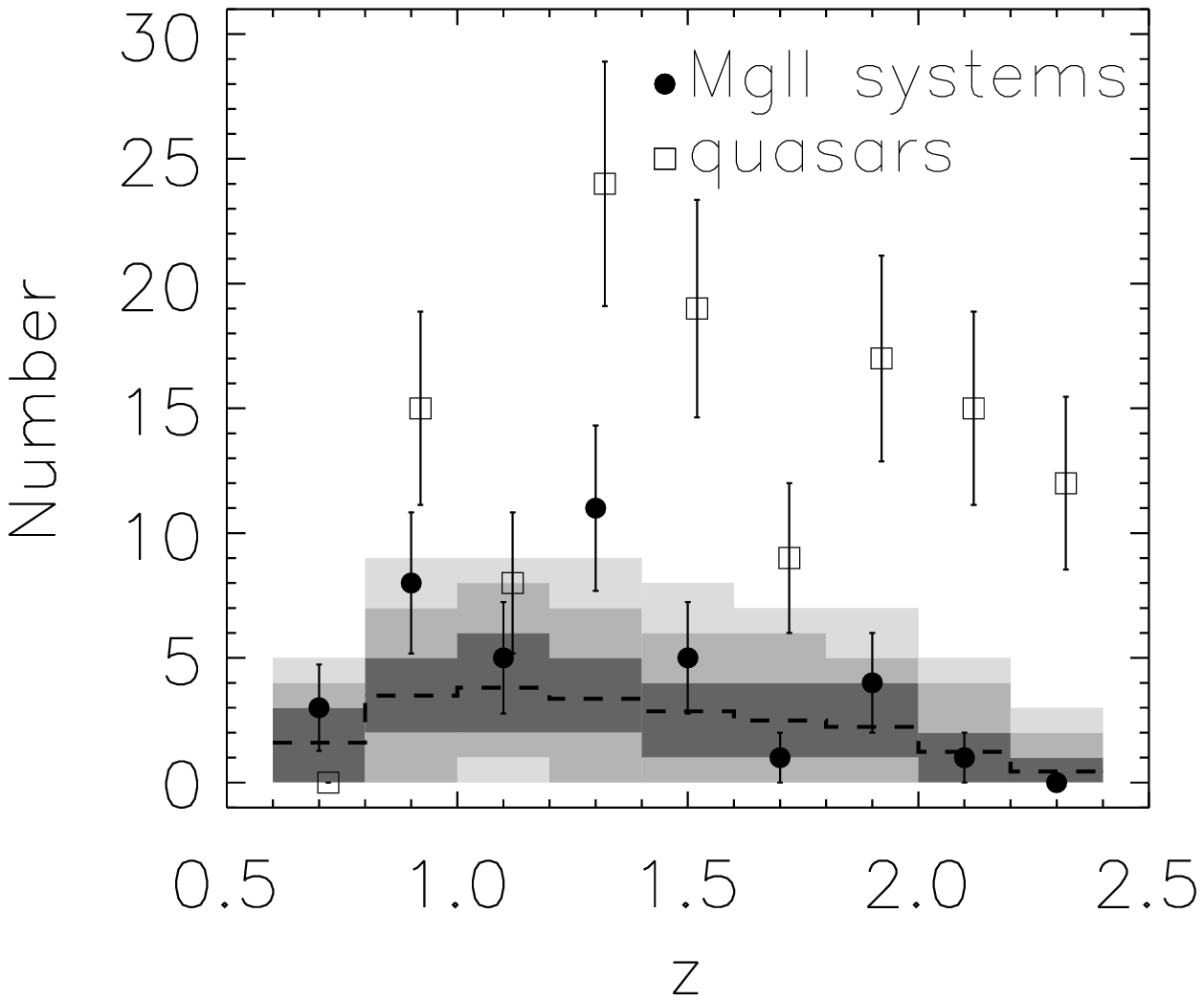} 
  \caption[]{(b) 
    }
\end{center}
\end{figure}

\clearpage
\begin{figure}[ht] 
\begin{center}
  \plotone{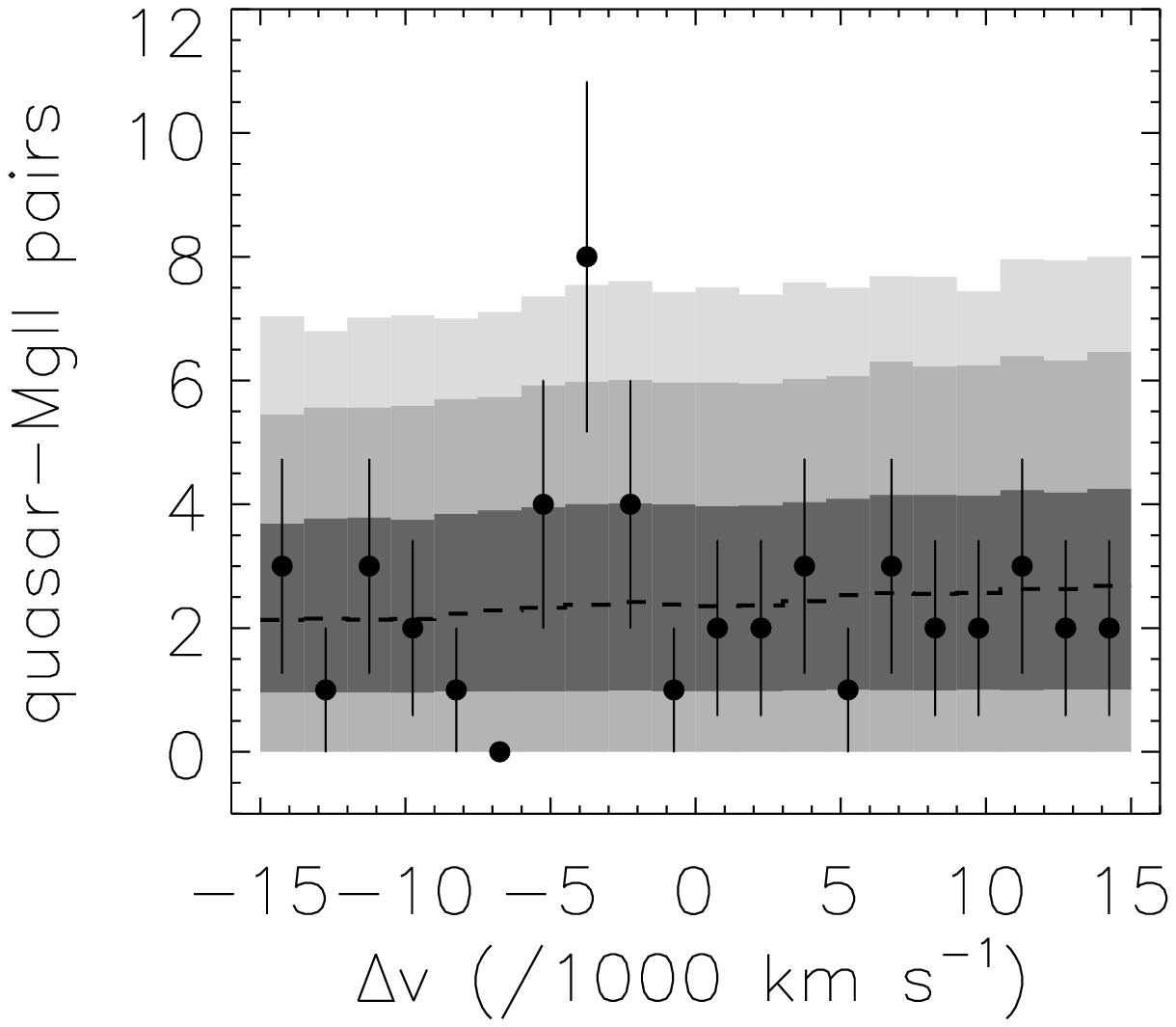} 
  \caption[]{Velocity distribution of quasar-\mgii\ pairs at
  $9h^{-1}$ Mpc projected separation.  Only pairs along 
  {\it different} lines of sight are counted.
  Negative velocity differences
  correspond to the quasar being at lower redshift than the \mgii\ absorber.
{\it Dashed
    line:}  mean expected number from 10000
  Monte Carlo simulations; {\it Shaded regions, darkest to lightest:} 
  68\%, 95\%, 99\% limits
  for the scatter about the expected mean;
  {\it Filled circles:} observed data, with error bars drawn assuming a
  Poissonian distribution as an illustration.  The errors are not exactly
  Poissonian, as any individual quasar or \mgii\ system could contribute to
  more than one pair.  However, as $9h^{-1}$ Mpc corresponds to 35 arcmin at
  $z=1.2$, there are relatively few pairs sharing quasars or \mgii\ absorbers,
  so the Poissonian approximation should give a reasonable error estimate.
  Note the overdensity at $-4500 < \Delta v < -3000$ \kms ,
  which has a random probability of occurrence of $P=0.002$.
    \label{fig:quasarmgiidndv}
    }
\end{center}
\end{figure}


\input{tab1.tex}
\input{tab2.tex}

\input{tab3.tex}

\begin{deluxetable}{cccc}
\tablecolumns{4}
\tablewidth{0pc} 
\tablecaption{\mgii\ sample
  \label{tab:absorbersample}
  }
\tablehead{
\colhead{sample} &
\colhead{total path $\Delta z$} &
\colhead{observed \mgii} &
\colhead{expected \mgii} \\
}
%
\startdata
Weak: $W_0(\lambda 2796)\geq 0.3$~\AA\  &      24.13       & 38              & 24.0 \\
Strong: $W_0(\lambda 2796)\geq 0.6$~\AA\  &      25.75       & 26              & 14.0 \\ 
\enddata 
\end{deluxetable}         

\input{tab5.tex}


%


\end{document}

%% file: tab1.tex

\begin{deluxetable}{ccccccccccr}
\tablecolumns{11}
\tablecaption{Target list / Observing log
  \label{tab:observinglog}
  }
\tablehead{
\colhead{quasar} &
\colhead{RA } &
\colhead{dec } &
\colhead{$z_Q$} &
\colhead{$U$} &
\colhead{$B$} &
\colhead{$V$} &
\colhead{$R$} &
\colhead{$I$} &
\colhead{$N^a$} &
\colhead{exp$^b$} \\
\colhead{} &
\colhead{J2000} &
\colhead{J2000} &
\colhead{} &
\colhead{} &
\colhead{} &
\colhead{} &
\colhead{} &
\colhead{} &
\colhead{} &
\colhead{(s)} \\
%
%
}
\startdata
%
J103937+0531    &    10:39:37.2 &05:31:46  &    2.035 &  18.20 & 19.13 & 19.17 & 18.88 & 18.11   &  5$^d$ & 9899 \\
J103952+0633    &    10:39:52.2 &06:33:22  &    1.395 &  18.14 & 18.90 & 18.73 & 18.19 & 18.13   &  4$^d$ & 5790 \\
J104007+0531    &    10:40:07.5 &06:25:09  &    2.383 &  18.05 & 18.54 & \nodata & \nodata & \nodata   &  1$^c$ & 2400 \\
J104117+0610    &    10:41:17.1 &06:10:17  &    1.273 &  15.42 & 17.03 & \nodata & \nodata & \nodata   &  1$^c$ &  900 \\
J104155+0612    &    10:41:55.7 &06:12:57  &    1.480 &  18.21 & 19.28 & \nodata & \nodata & \nodata   &  3$^d$ & 5800 \\
J104213+0628    &    10:42:13.5 &06:28:53  &    2.031 &  18.50 & 19.27 & 19.21 & 19.22 & 18.22   &  3$^d$ & 6600 \\
J104213+0619    &    10:42:13.6 &06:19:42  &    1.560 &  18.73 & 19.24 & 19.09 & 18.60 & 17.87   &  2$^d$ & 5700 \\
J104323+0422    &    10:43:23.6 &04:22:17  &    2.338 &  19.08 & 18.92 & 18.89 & 18.66 & 17.89   &  1$^c$ & 3000 \\
\enddata 
\tablenotetext{a}{number of exposures}
\tablenotetext{b}{total exposure time}
\tablenotetext{c}{observed during run 1 (1997 March 30 - April 1)}
\tablenotetext{d}{observed during run 1 and run 2 (1999 March 23-24)}
\end{deluxetable}         

%% file: tab2.tex
\begin{deluxetable}{ccrcrl}                                                                                                         
\tablewidth{0pc}                                                                                                                    
\tabletypesize{\footnotesize}                                                                                                       
\tablecolumns{6}                                                                                                                    
\tablecaption{Absorption line list for quasars                                                                                      
  \label{tab:linelist}                                                                                                              
  }                                                                                                                                 
\tablehead{                                                                                                                         
\colhead{Central wavelength } &                                                                                                     
\colhead{$\Delta \lambda^a $} &                                                                                                     
\colhead{Equivalent} &                                                                                                              
\colhead{$\Delta W$} &                                                                                                              
\colhead{signif$^b $} &                                                                                                             
\colhead{identification} \\                                                                                                         
  \colhead{$\lambda $ (\AA , vacuum)}               &                                                                               
  \colhead{}          &                                                                                                             
  \colhead{width $W$ (\AA )}       &                                                                                                
  \colhead{}     &                                                                                                                  
  \colhead{} &                                                                                                                      
  \colhead{/comments$^{c,d}$}\\                                                                                                     
%
}                                                                                                                                   
\startdata                                                                                                                          
\cutinhead{J103937+0531}                                                                                                            
5569.69 &      0.87 &	 1.543 &    0.232 &    6.651 &  \\                                                                            
5583.63 &      0.51 &	 2.014 &    0.221 &    9.113 &  \\                                                                            
5681.09 &      1.17 &	 1.709 &    0.249 &    6.863 &  \\                                                                            
5997.72 &      1.03 &	 0.733 &    0.218 &    3.362 & \mgii\  2796 $z=1.1448$ \\                                                     
6012.73 &      0.87 &	 0.471 &    0.177 &    2.661 & \mgii\  2803 $z=1.1448$ \\                                                     
6295.94 &      0.71 &	 1.112 &    0.228 &    4.877 & \mgii\  2796 $z=1.2515$? (bl)\\ 
6314.80 &      1.27 &	 0.455 &    0.167 &    2.725 & \mgii\  2803 $z=1.2515$? (bl)\\                                                
6358.44 &      0.51 &	 0.879 &    0.155 &    5.671 &  \\
\enddata                                                                                                                            
\tablenotetext{a}{$1 \sigma$ uncertainty in feature centroid}                                                                       
\tablenotetext{b}{significance ($W/\Delta W$)}                                                                                      
\tablenotetext{c}{bl = blend}                                                                                                       
\tablenotetext{d}{$\lambda_\oplus = $ in telluric absorption zone}                                                                          
\end{deluxetable}                                                                                                                   
%

%% file: tab3.tex
\begin{deluxetable}{clcccccrcrl}
\tabletypesize{\footnotesize}
\rotate
\tablecolumns{11}
\tablewidth{0pc} 
\tablecaption{\mgii\ systems 
  \label{tab:metalsys}
  }
\tablehead{
\colhead{quasar} &
\colhead{$z_{MgII}^a$} &
\colhead{type$^b$} &
\multicolumn{2}{c}{survey} &
\colhead{cl$^e$} &
\colhead{$W_0(\lambda 2796)$ } &
\colhead{$\sigma(\lambda 2796)$} &
\colhead{$W_0(\lambda 2803)$} &
\colhead{$\sigma(\lambda 2803)$} &
\colhead{comments} \\
\colhead{} &
\colhead{} &
\colhead{} &
\colhead{s$^c$} &
\colhead{w$^d$} &
\colhead{} &
\colhead{(\AA )} &
\colhead{} &
\colhead{(\AA )} &
\colhead{} &
\colhead{/other metals} \\
%
%
}
\startdata
%
J103937+0531 & 1.1448 & w & n & y & r & 0.342 $\pm$0.102 &  3.362 & 0.220 $\pm$0.083 &  2.661 & \\
J103937+0531 & 1.2515 & w & n & n & c & 0.494 $\pm$0.101 &  4.877 & 0.202 $\pm$0.074 &  2.725 & \\
J103937+0531 & 1.9748 & w & n & y & r & 0.436 $\pm$0.113 &  3.872 & 0.374 $\pm$0.083 &  4.492 & \\
J103952+0633 & 1.2459 & w & n & y & r & 0.329 $\pm$0.077 &  4.272 & 0.678 $\pm$0.081 &  8.409 & \feii\ $W_{0,2600}=0.246$\\
J103952+0633 & 1.3353 & w & n & y & r & 0.339 $\pm$0.085 &  4.000 & 0.311 $\pm$0.075 &  4.172 & \\
J104007+0531 & 1.3015 & w & n & y & r & 0.566 $\pm$0.068 &  8.293 & 0.426 $\pm$0.056 &  7.656 & \mgi ?; \feii\ $W_{0,2600}=0.295$ \\
J104007+0531 & 1.5706$^f$ & s & y & y & r & 3.462 $\pm$0.115 & 30.061 & 2.846 $\pm$0.101 & 28.142 & \feii\ $W_{0,2600}=1.305$ \\
J104155+0612 & 0.8829 & s & y & y & r & 0.933 $\pm$0.092 & 10.092 & 1.034 $\pm$0.088 & 11.794 & \\
\enddata 
%
\end{deluxetable}         

\setcounter{table}{2}
\begin{deluxetable}{clcccccrcrl}
\tabletypesize{\footnotesize}
\rotate
\tablecolumns{11}
\tablewidth{0pc} 
\tablecaption{\mgii\ systems (continued)
  \label{tab:metalsys2}
  }
\tablehead{
\colhead{quasar} &
\colhead{$z_{MgII}^a$} &
\colhead{type$^b$} &
\multicolumn{2}{c}{survey} &
\colhead{cl$^e$} &
\colhead{$W_0(\lambda 2796)$ } &
\colhead{$\sigma(\lambda 2796)$} &
\colhead{$W_0(\lambda 2803)$} &
\colhead{$\sigma(\lambda 2803)$} &
\colhead{comments} \\
\colhead{} &
\colhead{} &
\colhead{} &
\colhead{s$^c$} &
\colhead{w$^d$} &
\colhead{} &
\colhead{(\AA )} &
\colhead{} &
\colhead{(\AA )} &
\colhead{} &
\colhead{/other metals} \\
%
%
}
\startdata
J104357+0438 & 0.8734 & v & n & n & c & 0.147 $\pm$0.045 &  3.235 & 0.476 $\pm$0.048 & 10.011 & \\
J104357+0438 & 1.0255$^f$ & s & y & y & r & 0.684 $\pm$0.062 & 11.080 & 0.523 $\pm$0.059 &  8.825 & \mgi ,  \\
             &        &   &   &   &   &                  &        &                  &        & \feii\ (bl) $W_{0,2600}=0.604$ \\
J104357+0438 & 1.5702 & s & y & y & r & 0.757 $\pm$0.056 & 13.608 & 0.500 $\pm$0.057 &  8.795 & \\
J104409+0531 & 1.0710 & w & n & y & r & 0.362 $\pm$0.075 &  4.839 & 0.710 $\pm$0.088 &  8.077 & \\
J104529+0623 & 0.7726 & s & y & y & r & 0.739 $\pm$0.155 &  4.781 & 0.625 $\pm$0.145 &  4.307 & \\
J104529+0623 & 1.0769 & w & n & y & r & 0.491 $\pm$0.085 &  5.757 & 0.281 $\pm$0.083 &  3.370 & \\
J104529+0623 & 1.1192 & v & n & n & d & 0.270 $\pm$0.059 &  4.548 & 0.374 $\pm$0.063 &  5.962 & \\
\enddata 
\end{deluxetable}         
\setcounter{table}{2}
\begin{deluxetable}{clcccccrcrl}
\tabletypesize{\footnotesize}
\rotate
\tablecolumns{11}
\tablewidth{0pc} 
\tablecaption{\mgii\ systems (continued)
  \label{tab:metalsys3}
  }
\tablehead{
\colhead{quasar} &
\colhead{$z_{MgII}^a$} &
\colhead{type$^b$} &
\multicolumn{2}{c}{survey} &
\colhead{cl$^e$} &
\colhead{$W_0(\lambda 2796)$ } &
\colhead{$\sigma(\lambda 2796)$} &
\colhead{$W_0(\lambda 2803)$} &
\colhead{$\sigma(\lambda 2803)$} &
\colhead{comments} \\
\colhead{} &
\colhead{} &
\colhead{} &
\colhead{s$^c$} &
\colhead{w$^d$} &
\colhead{} &
\colhead{(\AA )} &
\colhead{} &
\colhead{(\AA )} &
\colhead{} &
\colhead{/other metals} \\
%
%
}
\startdata
J104747+0456 & 0.8341 & s & y & y & r & 1.887 $\pm$0.112 & 16.806 & 0.238 $\pm$0.063 &  3.767 & \\
J104747+0456 & 1.2455 & w & n & n & c & 0.596 $\pm$0.100 &  5.951 & 0.180 $\pm$0.061 &  2.949 & \\
J104747+0456 & 1.9312 & s & y & y & r & 0.915 $\pm$0.082 & 11.175 & 0.229 $\pm$0.059 &  3.851 & \\
J104747+0456 & 2.0728$^f$ & s & a & a & r & 2.360 $\pm$0.071 & 33.271 & 2.398 $\pm$0.114 & 21.057 & \feii\ $W_{0,2600}=1.109$ , \mgi ; \civ ?  \\
J104840+0535 & 0.9882$^f$ & s & y & y & r & 1.118 $\pm$0.071 & 15.648 & 1.510 $\pm$0.071 & 21.148 & \feii\ $W_{0,2600}=0.677$ \\
J104840+0535 & 1.2460 & w & n & n & c & 0.318 $\pm$0.067 &  4.735 & 0.414 $\pm$0.079 &  5.219 & \\
J104840+0535 & 1.2728 & w & n & y & r & 0.517 $\pm$0.062 &  8.333 & 0.198 $\pm$0.051 &  3.913 & \feii\ $W_{0,2600}=0.227$? \\
J104914+0414 & 1.2458 & w & n & y & r & 0.428 $\pm$0.086 &  4.959 & 0.559 $\pm$0.072 &  7.801 & \\
\enddata 
\tablenotetext{a}{redshift is determined by centroid of both \mgii\ components, with an attempt to correct for blending.}
\tablenotetext{b}{s: strong ($W_{0,\lambda 2796}\geq 0.6$ \AA ); w: weak ($0.3\leq W_0 < 0.6$\AA ); v: very weak ($W_0<0.3$ \AA )}
\tablenotetext{c}{y: in strong survey; n: not in strong survey; a: associated system (excluded)}
\tablenotetext{d}{y: in weak survey; n: not in weak survey; a: associated system (excluded)}
\tablenotetext{e}{class: r: real; c: candidate; d: doubtful}
\tablenotetext{f}{candidate damped system ($W_{0,MgII\, 2796}\geq 0.5$ \AA\ and $W_{0,FeII\, 2600}\geq 0.5$ \AA )}
\end{deluxetable}

%% file: tab5.tex
\begin{deluxetable}{lcccccc}
\tablecolumns{7}
\tablewidth{0pc} 
\tablecaption{Quasar-\mgii\ pairs with $-4500 < \Delta v < -3000$ \kms\ and $\Delta \ell <9h^{-1}$ Mpc
  \label{tab:quasarmgiipairs}
  }
\tablehead{
\colhead{quasar} &
\colhead{$z_Q$} &
\colhead{\mgii\ toward} &
\colhead{$z_{MgII}$} &
\colhead{$\Delta v^a$} &
\colhead{$\Delta \ell^b$} &
\colhead{class } \\
\colhead{} &
\colhead{} &
\colhead{quasar} &
\colhead{} &
\colhead{} &
\colhead{($h^{-1}$ Mpc)} &
\colhead{(strong/weak)} \\
}
%
\startdata
J103908+0459	 &  1.942   &   J103936+0531	&      1.9748  &   -3330     &    8.1 &  w  \\   
J104117+0610	 &  1.273   &   J104007+0531	&      1.3015  &   -3740     &    5.9 &  w  \\   
J104149+0643	 &  1.316   &   J104155+0612	&      1.3460  &   -3860     &    8.0 &  w  \\   
J104155+0612	 &  1.480   &   J104213+0628	&      1.5072  &   -3270     &    4.3 &  s  \\   
J104200+0441	 &  1.97\phantom{0}&J104323+0422&      2.0121  &   -4220     &    7.0 &  s  \\   
J104336+0558$^c$ &  1.236   &   J104155+0612	&      1.2628  &   -3570     &    7.4 &  w  \\   
J105010+0432	 &  1.217   &   J104914+0414	&      1.2458  &   -3870     &    5.9 &  w  \\   
J105030+0430	 &  1.215   &   J104914+0414	&      1.2458  &   -4140     &    6.4 &  w  \\   
\enddata 
\tablenotetext{a}{$\Delta v \equiv c(z_Q-z_{MgII})/(0.5(2+z_Q+z_{MgII}))$}
\tablenotetext{b}{projected distance in the plane of the sky}
\tablenotetext{c}{position not in Veron-Cetty \& Veron (2001); RA=10:43:36.6,  dec=05:58:48 (J2000)}
\end{deluxetable}